# PERFORMANCE ANALYSIS FOR THE NEW G-2 EXPERIMENT*

D. Stratakis[†], M. E. Convery, C. Johnstone, J. Johnstone, J. P. Morgan, M. J. Syphers[1]
Fermi National Accelerator Laboratory, Batavia IL, USA
J. D. Crmkovic, W. M. Morse, V. Tishchenko, Brookhaven National Laboratory, Upton NY, USA
N. S. Froemming, University of Washington, Seattle WA, USA
M. Korostelev, University of Lancaster/Cockcroft Institute, UK
[1]also at Northern Illinois University, DeKalb IL, USA

*Abstract*

The new g-2 experiment at Fermilab aims to measure the muon anomalous magnetic moment by a fourfold improvement in precision compared to the BNL experiment. Achieving this goal requires the delivery of highly polarized 3.094 GeV/c muons with a narrow ±0.5% $\Delta p/p$ acceptance to the storage ring. In this study, we describe a muon capture and transport scheme that should meet this requirement. First, we present the conceptual design of our proposed scheme wherein we describe its basic features. Then, we detail our numerical model and present a complete end-to-end simulation of all g-2 beamlines.

## INTRODUCTION

The Muon g-2 Experiment, at Fermilab [1], will measure the muon anomalous magnetic moment, $\alpha_\mu$ to unprecedented precision: 0.14 parts per million. The worth of such an undertaking is coupled to the fact that the Standard Model prediction for $\alpha_\mu$ can also be determined to similar precision. Thus, the comparison between experiment and theory provides one of the most sensitive tests of the completeness of the model.

Most of the new experiment improvements will be based on increased statistics. Thus, it is essential to maximize the transmission of polarized muons within the acceptance of the g-2 storage ring. The present baseline lattice, however, has a series of bending sections, elevation changes as well as complex injection and extraction schemes. Thus, the goal of this paper is to develop a simulation model for the g-2 lattice and evaluate numerically its performance. We initiate our simulation at g-2 target and incorporate G4Beamline [2], a code that includes basic physical processes such as spin tracking and muon decay.

## BEAMLINE DESCRIPTION

A layout of the Fermilab Muon Campus is shown in Fig. 1. A 8 GeV kinetic energy proton beam impinges a cylindrical core nickel alloy target surrounded by a thick casing of beryllium. Pions produced in the target will be focused by a Li lens and subsequently transported into the M2 line via a $3^0$ pulsed bending magnet which is designed to select positive particles with a momentum around 3.1 GeV/c.

M2 line starts with a four matching quadrupole triplet, followed by eight more quadrupoles and a dipole, which match into the lattice of the M3 line. While peak beta functions occur in the triplet, further downstream they vary between 2 to 25 m so that to maximize muon capture [see Fig 2(a)]. For (g -2), the M2 line continues across the tunnel, intersecting the M3 line 50 m further downstream from the Li lens. At the intersection, a large aperture dipole provides the second $3^0$ bend for the beam to match the M3-line trajectory. Further downstream, an $18.5^0$ horizontal bend is provided by a specialized insertion created by a two dipole achromatic bend. The beam continues for 63.0 m to the beginning of the geometric and optical matching section between the M3 line and the Delivery Ring (DR) injection point in the D30 straight section. The beam is injected vertically with a bend produced by a combination of a C-magnet, followed by a large-aperture focusing quad Q303 and a pulsed magnetic septum dipole in the DR (Fig. 3). Two kicker modules downstream of Q301 (not shown in Fig. 3) close the trajectory onto the orbit of the DR. The length from the face of the Li lens to the DR entrance is 280 m.

The DR is a 505 m rounded triangle and is divided into 6 sectors numbered 10-60. Each sector contains 19 quadrupoles and 11 dipoles. There are three straight sections wherein the dispersion is low while the arcs are dispersive regions. A typical cell in the arcs is comprised of an F-quadrupole with similarly oriented sextupoles on either side followed by a dipole or drift region, then a D-quadrupole with a similar arrangement.

For extraction, the beam is kicked horizontally via two kickers and then, a Lambertson and C-magnet pair will be used, in conjunction with the intervening D2Q5 quadrupole, to bend the beam upward out of the DR. Then, the beam passes through a series of vertical steering magnets through part of the M4 line, then bends upward into the 100 m long M5 line and continues to the (g-2) Storage Ring. At the end of M5, a final focus section is designed to provide optical matching to the storage ring.

## SIMULATION DETAILS

Figure 2(b) displays the simulated performance along the M2-M3 lines. The black solid and dashed curves show the number of pions with and without decays, respectively. The theoretical calculated number of pions at the channel end agrees well with the simulation. At the ring entrance the number of pions overpass the muons by one order of magnitude.

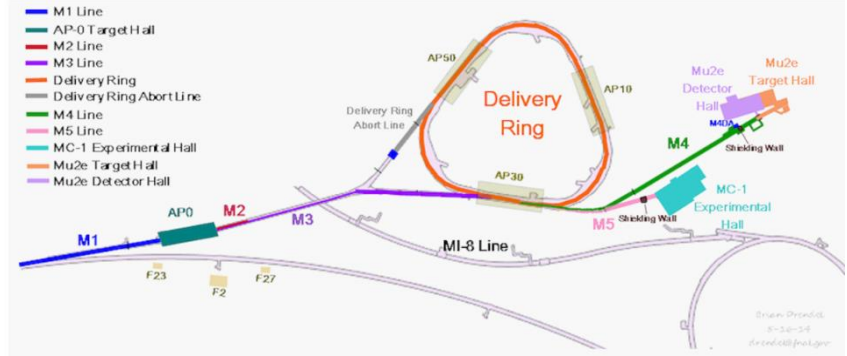

Figure 2: Example of a full-width figure showing the JACoW Team at their annual meeting in 2015. This figure has a multi-line caption that has to be justified rather than centred.

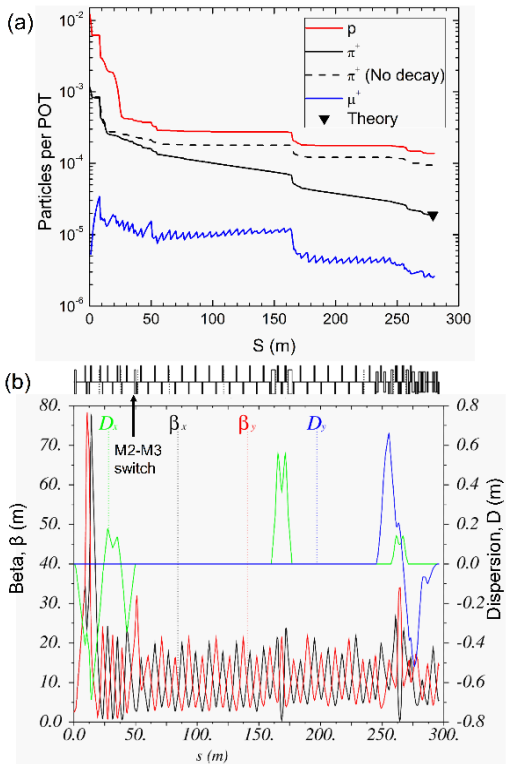

Figure 2: (a) Optics model and, (b) simulated performance along the M2-M3 beamlines up to the DR entrance.

The loss of particles near 160 m is due to the $18.5^0$ horizontal bend. The protons entering the DR are removed through a single kicker magnet in Sector 50 [1, 5]. Figure 4 displays the number of secondary particles vs the number of turns in the DR. After the second turn all pions have decayed while the transmission after the four turns remains near 90%. Figure 5(a) displays the simulated performance along the M4-M5 lines. The simulation ends 1.59 m after the last final focus magnet and upstream of the inflector. The number of muons per proton on target (POT) within the ring acceptance $\Delta p/p = \pm 0.5\%$ is $2.4\times10^{-7}$ which is in agreement with Ref. 1. The final beam is centered at 3.091 GeV/c with a spread $\Delta p/p = 1.2\%$. Systematic effects on the measurement of $a_\mu$ occur when the beam has a correlation between the muon spin direction and its momentum [1]. Due to anomaly, the muon spin precesses relative to muon momentum, in the DR, by an angle $\varphi_a = 2\pi N\gamma a_\mu$, where $N$ is the turn number and $\gamma$ is the relativistic factor. The slope of the spin-momentum correlation is then $d\varphi_a/dp = 2\pi N a_\mu/m_\mu$ where $m_\mu$ is the muon rest mass. The correlation between spin precession angle and momentum of the muon at $N$=4 is shown in Fig. 6. Linear fit to the data gives a slope of 0.301 mrad/MeV/c while the one predicted by the theory is 0.278 mrad/MeV/c.

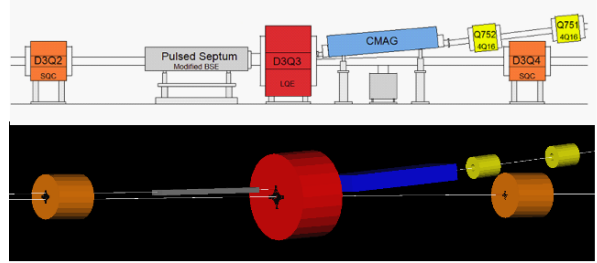

Figure 3: Conceptual design and simulation model of the injection line. Similar idea is used for extraction.

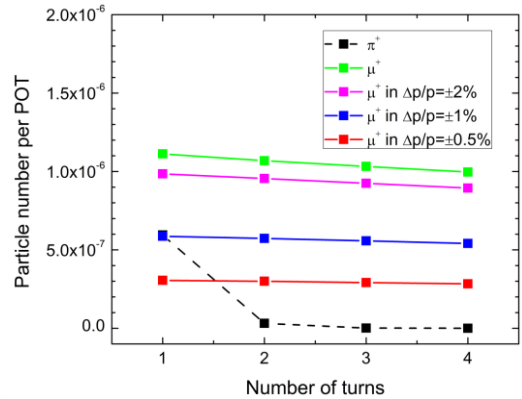

Figure 4: Particles per POT vs. turn number along the DR.

## STORAGE-RING SIMULATIONS

When the muon beam is injected into the g-2 storage ring, it must first pass through a superconducting septum magnet designed to facilitate injection by cancelling the main field of the dipole storage-ring magnet. The muon beam is then "kicked" onto the proper storage orbit by pulsed electromagnets. Once inside the storage region, vertical focusing is provided by electrostatic quadrupoles to prevent the beam from diverging out of the ring. A Geant-based simulation has been written to model muon injection into the *g*-2 storage ring, beam dynamics, and detector response (Fig. 7).

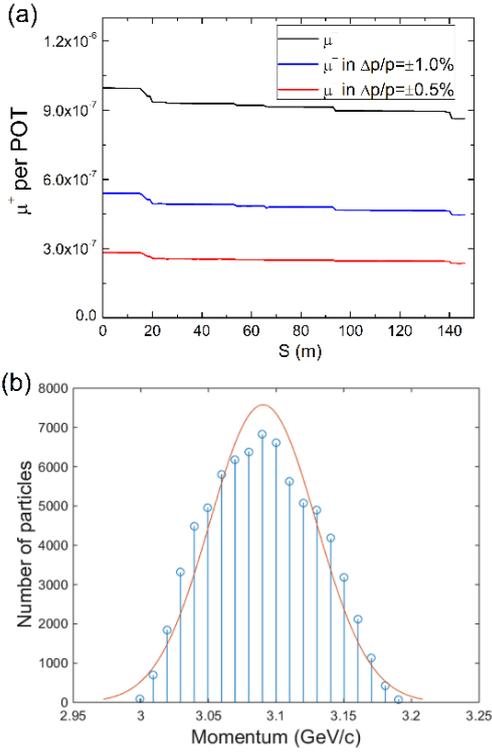

Figure 5: Performance along the M4-M5 beamlines, and (b) momentum distribution at the inflector entrance.

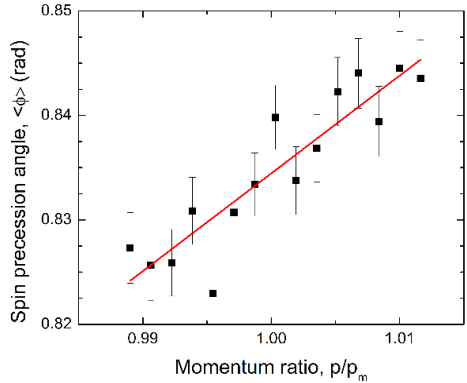

Figure 6: Spin-momentum correlations after the fourth turn in the DR. Note that $p_m$ is the muon magic momentum.

One improvement in E989 will be a system of 3 detectors to continually monitor the muon beam as it passes from the final focus into the storage ring. In this region, the muon beam is focused to a waist in order to pass through the narrow aperture ($\Delta x=\pm 9$mm, $\Delta y=\pm 28$mm) of the 1.7 m-long superconducting septum magnet without scraping and beam loss occurring. Simulations indicate that, although the horizontal dispersion of the storage ring is 8.6m (at a quad field index of $n=0.175$), the injection tune actually prefers *zero* horizontal dispersion due to the narrow aperture of the septum magnet.

Example injection tunes and corresponding muon storage rates are shown in Fig. 8(a). The beam dynamics of the stored muon distribution reflects not only the injection tune, but also how the beam was kicked onto the storage orbit. An example is shown in Fig. 8(b). Here, the oscillation of the beam centroid indicates the kicker strength/timing could be further optimized. All in all, major strides have been made in muon *g*-2 simulations, which have had a real, positive impact on the experiment overall.

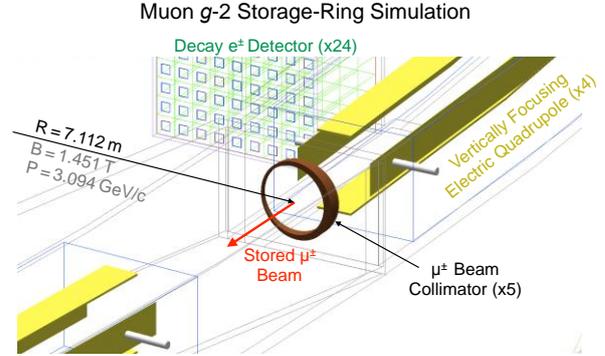

Figure 7: Geant-based simulation model of the muon *g*-2 storage ring.

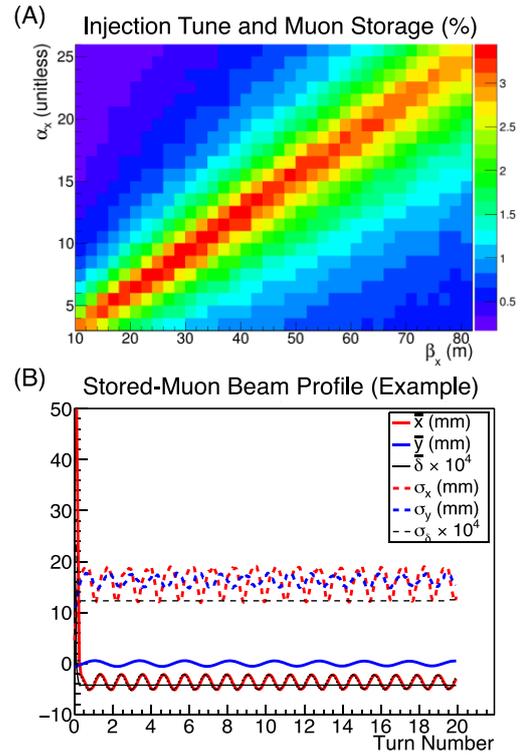

Figure 8: (A) Example Twiss parameters at injection and corresponding muon storage. Injection is most sensitive to horizontal Twiss parameters since the narrowest aperture is $\Delta x=\pm 9$mm. (B) Example beam profile of stored muons.